**Tabakov Valery**

*Ph.D of technical sciences, assistant professor, Private Joint Stock Company "Higher education institution "Interregional Academy of Personnel Management"", Kiev city, Ukraine*

# Integration of higher IT education in Ukraine in the global IT-educational space.

**Abstract:** The article presents the results of a study of the current state of higher IT education system in Ukraine. The problems of reforming higher IT education system of Ukraine in accordance with the commitments made by Ukraine in connection with the ratification of the EU-Ukraine Agreement Law of Ukraine № 1678-VII of September 16, 2014. An indicator of the presence or absence of a real reform of the system of higher IT education in Ukraine is detected. A comparative analysis of lists of IT-specialties of higher education in Ukraine in 2005 and 2015 with similar lists adopted by the international system of higher IT education is made. A discrepancy between the list of IT-specialties in Ukraine and international list of IT specialties are identified. The conclusion about the need for immediate correction of the list of higher education in Ukraine IT-specialties in order to bring it into line with international standards. It recommended a series of actions that will lead to the solution of the problem.

## 1. Introduction

High quality of higher education in Ukraine is the aim of the Ukrainian society and the key to its success. Movement of Ukrainian educational community in this direction is due to the implementation of the obligations imposed on Ukraine by the EU-Ukraine agreement, which was ratified by the Law of Ukraine № 1678-VII dated from 16 September 2014 [1]. Article 431 of the Agreement specifically defined goal of enhancing EU-Ukraine cooperation in the field of higher education, namely the reform and modernization of higher education systems; to promote convergence in the field of higher education that takes place in the framework of the Bologna process; improving the quality of higher education and the importance of deepening cooperation between higher education institutions; empowerment of institutions of higher education; activation of the mobility of students and teachers; and facilitate access to higher education.

With regard to the reforming of the higher education system of Ukraine, this paragraph of Article 431 of the Agreement is implemented in accordance with the new Law of Ukraine "On Higher Education", № 1556-VII of the July 1, 2014 [2], and the problem is that all the norms of this law were implemented. Ukrainian realities are such that no matter how wonderful was the Law of Ukraine, today it is no guarantee of its implementation. So the question of Bologna processing of higher education, ensuring its quality, mobility of students and teachers today as they were 10 years ago, come across to stiff resistance of outdated Ukrainian governing system. It can be stated that today's "reform" of higher education in Ukraine are formal and are not changes in the system of higher education in Ukraine on the merits.

One of the indicators of the state of the governing system of the Ukrainian higher education is a bureaucratic sabotage of section V of the Law of Ukraine "On Higher Education", which is called "Ensuring the quality of higher education." The management of quality of higher education is assigned by Law on the National Agency for Quality Assurance in Higher Education. The realities of the Ukrainian educational system management are such that the newly elected National Agency for Quality Assurance in Higher Education cannot take up its duties, which was to begin on September 1, 2015, due to the lack of agreement regarding its personal structure by the former Minister of Education and Science of Ukraine Sergei Kvit. As a result, the processes of licensing and accreditation of educational programs of higher educational institutions in Ukraine continued to be carried out on the old "corruption" procedures.

Delaying of reforms of higher education in Ukraine is particularly painfully felt by the education sector in the IT sector, which, in comparison with other areas and sectors of higher education is the most dynamic. Irreparable loss of the pace of reforms of education accurately characterizes 9 and 14 commandments of Harvard: "Time flies" and "even now, your enemies are already leafing through the book." Therefore, the relevance of research of problems of intensifying of Ukraine-EU Cooperation in Higher IT education field and integration of higher IT education in global IT-educational environment is undeniable.

Our research aims to propose an Action Plan, the implementation of which will lead to a real integration of IT sector of higher education of Ukraine in global IT- educational space.

## 2. Methods for Evaluation and Analysis

To achieve this, we have made an analysis of current Ukrainian and international legal frameworks governing the education and qualification process in the IT sector of higher education, revealed the difference between the Ukrainian and international global standards, and identified measures to modernize the Ukrainian IT sector of higher education, the implementation of which will bring standards and realities of the Ukrainian IT education to the world standards and lead to the practical implementation of the provisions of Article 431 of the Agreement Ukraine-EU on sector of higher IT education of Ukraine, namely the convergence in higher IT education that takes place in the framework of the Bologna process, to deepen cooperation between the higher education institutions engaged in training specialists with higher education in IT, and for increased mobility of students and teachers in the IT sector of higher education, as a result of the harmonization of Ukrainian and world IT higher education sector.

## 3. Results

Firstly, we have defined a list of IT specialties of Ukrainian universities in force until September 1, 2015 (Resolution of CMU from 13.12.2006, № 1719) [3]. This list is quite diverse and includes 1) Engineering (direction) field of study in the fields of Computer Science and Engineering: computer science, software engineering, and computer engineering; 2) Engineering specialty of industry in the field of automation and control: systems engineering and automation and computer-integrated technologies; 3) natural science specialty, the field of system science and cybernetics: applied mathematics, computer science and systems analysis; 4) specialty group of information security section of the security: the security of information and communication systems, systems of technical protection of information and management of information security, and 5) specialties from of humanities and art in the field of economics - economic, cybernetics. In general, the twelve directions. Of these, the purely IT-related are just three: computer science, software engineering, and computer engineering.

Secondly, we analyzed the new list of areas of knowledge and specialties, which are trained candidates of higher education, approved by the Decree of the Cabinet of Ukraine № 266 dated 29 April 2015, which entered into force on 1st September 2015 [4]. According to the list five disciplines in Ukraine are considered to be IT-related: software engineering, computer science and information technology, computer engineering, systems analysis and cyber security. To IT list approved by the Decree of the Cabinet of Ministers, in our opinion, may be implicitly included only four majors, and the fifth - the system analysis, got into the list by mistake, and only at the personal request of the rector of the National Technical University of Ukraine "KPI" Michael Zgurovsky.

The development of the world's highest standards of IT education in 2005 led to the publication of Computing Curricula 2005 [5], which is a de facto approved international list of IT-related specialties as computer engineering, computer science, information systems, information technology and software engineering. Regarding international list of IT-related specialties, we conclude that the state of the Ukrainian system of higher education in the IT sector as well as the system of qualifications of Ukrainian specialists with higher IT education, corresponds to the world standards of management systems of IT education and qualifications system of IT specialists ten years ago, and even then not completely, but partially. About the approximation of IT sectors of higher education, declared in Article 341 of the Agreement EU-Ukraine, today it is possible to ascertain, as partly achieved and with a lag of Ukraine from the European Union for ten years.

The current state of the world's IT standards of the higher education was recorded two years ago in the document called Computing Science Curricula 2013 [6], which summarizes the achievements of the world higher education in IT sphere. According to this document the world nomenclature of specialties of higher IT education consists not even of five specialties, as it was in 2005, and only one specialty - Computer science. This specialty is included in the list of specialties approved by the Resolution of the Cabinet of Ministers of Ukraine № 266 dated 29 April 2015, but not as a separate generalized name corresponding to the modern international system of qualifications of IT specialists with higher education, but as one of the five specialties, the names of which are reminiscent of the names of IT specialties decade ago. Why does the Ministry of Education and Science of Ukraine, together with the Cabinet of Ministers of Ukraine create the conditions in which the name of the specialties of Ukrainian IT specialists will not match to the current international system of qualifications, one can only guess. But it is clear that such actions of the Ukrainian managers are the direct sabotage of the commitments undertaken by Ukraine on 16 September 2014 before the European Community.

An example of the application of the current international system of qualifications of IT-specialists is the Guidelines for Bachelor of Computer Science of the University of Colorado 2014-2015 [7]. Said guidance provides the preparation of IT specialists with higher education in the same direction as of Computer Science, which identified the following fundamental components of the curriculum such as Informatics 1 (programming), Informatics 2 (data structures), Computer systems, Algorithms and Programming languages principles. The rest are additional disciplines, specializing the students in areas such as self-determination, computational biology, human-centering (automated) calculation, network devices and systems, Software Engineering, Systems.

### 4. Conclusions and Future Work

1. The state of management system of higher IT education in Ukraine can be characterized as below the level of the global management system for higher IT education for ten years.

2. The system of qualifications and specializations of IT specialists of Ukraine does not correspond to the modern world IT specialties nomenclature defined in Computing Science Curricula in 2013.

3. The list of IT specialties, approved by the Decree of the Cabinet of Ministers of Ukraine № 266 of 29 April 2015, constitutes an obstacle to the implementation of Article 431 of the Association Agreement EU-Ukraine.

4. The urgent task of Ukrainian IT-community is to bring the range of IT professionals whose training is carried out by the educational institutions of Ukraine, in accordance with the nomenclature of the global IT specialties.